# Integrating Functional Oxides with Graphene

X. Hong[1], K. Zou[2], A. M. DaSilva[2], C. H. Ahn[4], and J. Zhu[2,3]

1. Department of Physics and Astronomy and Nebraska Center for Materials and Nanoscience, University of Nebraska-Lincoln, Lincoln, Nebraska, 68588, USA
2. Department of Physics, The Pennsylvania State University, University Park, Pennsylvania, 16802, USA
3. The Materials Research Institute, The Pennsylvania State University, University Park, Pennsylvania, 16802, USA
4. Department of Applied Physics, Yale University, New Haven, Connecticut, 06520-8284, USA

**Abstract**

Graphene-oxide hybrid structures offer the opportunity to combine the versatile functionalities of oxides with the excellent electronic transport in graphene. Understanding and controlling how the dielectric environment affects the intrinsic properties of graphene is also critical to fundamental studies and technological development of graphene. Here we review our recent effort on understanding the transport properties of graphene interfaced with ferroelectric $Pb(Zr,Ti)O_3$ (PZT) and high-$\kappa$ $HfO_2$. Graphene field effect devices prepared on high-quality single crystal PZT substrates exhibit up to tenfold increases in mobility compared to $SiO_2$-gated devices. An unusual and robust resistance hysteresis is observed in these samples, which is attributed to the complex surface chemistry of the ferroelectric. Surface polar optical phonons of oxides in graphene transistors play an important role in the device performance. We review their effects on mobility and the high source-drain bias saturation current of graphene, which are crucial for developing graphene-based room temperature high-speed amplifiers. Oxides also introduce scattering sources that limit the low temperature electron mobility in graphene. We present a comprehensive study of the transport and quantum scattering times to differentiate various scattering scenarios and quantitatively evaluate the density and distribution of charged impurities and the effect of dielectric screening. Our results can facilitate the design of multifunctional nano-devices utilizing graphene-oxide hybrid structures.

1.  **Introduction**

The unusual electronic properties and robust chemical and mechanical properties of graphene make it a promising material base for developing nano-electronic and spintronic applications [1-3]. In the past few years, graphene-based high performance prototype devices have been demonstrated and new device concepts have been developed, including radio-frequency transistors operating at 100 GHz [4], graphene nanoribbon transistors [5], spin valve devices with long spin coherent lengths [6, 7], and *pn* junctions and electron lensing devices based on Klein tunneling [8, 9]. An essential element for building up logic and memory operations in device applications is the dielectric gate. While existing device designs predominantly utilizes $SiO_2$ as the gate material, interfacing graphene with various functional oxides may offer tremendous new opportunities.

Oxide materials exhibit a wide range of electronic and magnetic properties, including high dielectric constant (high-$\kappa$), ferroelectricity, magnetism, and superconductivity [10]. Integrating the versatile functionalities of oxides with graphene can significantly broaden the spectrum of graphene applications, ranging from high efficiency field effect gating and local density modulation to nonvolatile memory and spintronic devices. One challenge, however, is that graphene is only a single atomic layer thick and is highly influenced by its environment. The intrinsic mobility of graphene set by longitudinal acoustic (LA) phonon scattering can reach $\sim 10^5$ $cm^2$/Vs at room temperature [11], while charge traps and remote surface optical phonons in adjacent dielectric layers can significantly reduce the mobility [12]. For example, graphene fabricated on the widely used $SiO_2$ substrates has a mobility ceiling of 20,000 $cm^2$/Vs [12, 13], one order of magnitude lower than those reported on suspended graphene [14, 15]. For graphene prepared on substrates, high mobility close to the intrinsic LA limit has only been observed on those utilizing single crystal substrates, such as Pb(Zr, Ti)$O_3$ (PZT) [16, 17] and Boron Nitride (BN) [18], further highlighting the important role played by the dielectric surface/interface properties in limiting graphene's performance. Understanding and controlling how the dielectrics affect the electron transport in graphene is thus critical to achieving the full fundamental and technological potential of graphene.

In this review paper, we will discuss the opportunities and impacts that various functional oxide materials bring to graphene from three aspects. The first topic is the unusual transport properties of graphene integrated with single crystal ferroelectric oxides. Ferroelectrics exhibit a spontaneous polarization that is switchable via an electric field larger than the coercive field of the material. The polarization of the ferroelectric oxide PZT can reach 50 $\mu C/cm^2$, corresponding to a two dimensional (2D) carrier density of $3 \times 10^{14}/cm^2$, more than one order of magnitude higher than what can be induced through conventional $SiO_2$ gates. Utilizing this bi-stable polarization field, one can achieve reversible switching of the carrier density and resistivity in graphene, which forms the foundations of nonvolatile memory operations. The ferroelectric field effect has previously been explored in nanowires and carbon nanotubes [19, 20], and has recently been applied to graphene [16, 17, 21, 22]. In addition, crystalline PZT thin films have well defined surface states and high dielectric constant compared to amorphous $SiO_2$, which can reduce scattering from interfacial charged impurities and provide effective dielectric screening. On graphene devices interfaced with single crystal PZT thin films, we have observed superb carrier mobility and unusual resistance hysteresis, which have been correlated with the dielectric and surface properties of the ferroelectrics [16, 17].

The second topic of interest is the impact of remote surface optical (RSO) phonons of the gate oxide layer on the performance of graphene transistors. Scattering from these phonon modes plays a major role in limiting high temperature mobility and the saturation current [23], which are important parameters that affect the operation temperature and speed for electronic devices. Materials of particular interest are the high-$\kappa$ oxides, which have soft phonon modes that are fully excited at room temperature [24-28]. High-$\kappa$ oxides have been widely utilized in both fundamental studies and technological development. For example, as current Si-based electronic devices are rapidly approaching the fundamental scaling limits, $HfO_2$ has been employed in transistors to achieve lower operation voltage and static power consumption. Due to their high dielectric constant, these materials can effectively screen the charged scatterers and improve low temperature mobility [29]. Since they can be deposited relatively easily on graphene without degrading graphene's intrinsic properties, these oxides have also been used as the top-gate in a double-gate structure to achieve local density modulation, a control that is not available through a global gate substrate. This double-gate structure has been used to explore a range of novel phenomena, including bandgap control in bilayer graphene [30, 31] and introducing a tunable energy barrier for Klein tunneling [8, 9]. Given the high research and application potential of graphene-high-$\kappa$ oxide hybrid structures, it is important to examine the impact of such integration on the intrinsic properties of graphene. We have investigated both theoretically and experimentally the effect of the RSO phonon from $SiO_2$ and high-$\kappa$ oxides on graphene's mobility [27] and saturation current [26], which are crucial for developing room temperature high-speed devices.

In the last section, we describe a set of experiments to probe the nature of the scattering sources in graphene devices. These studies utilize a sensitive transport probe, the ratio of the transport and quantum scattering times, to reveal important characteristics of the scattering sources [32-34]. We apply this method to evaluate quantitatively the type, density, and location of the scattering sources and the effect of dielectric screening on charged impurities in graphene [33].

## 2. Materials Preparation

For PZT-gated $n$-layer graphene ($n$-LG) devices, we use 200-400 nm thick single crystal Pb(Zr$_{0.2}$Ti$_{0.8}$)O$_3$ (PZT) films epitaxially grown on Nb-doped (001) SrTiO$_3$ (STO) substrates [16, 17]. The PZT films show high crystallinity and 3-4 Å surface roughness, with the as-grown polarization pointing uniformly towards the substrate. A negative voltage larger than the coercive voltage (-6 to -8 V for these films) relative to the substrate is required to reverse the polarization [17]. Limited by gate leakage, the polarization state of PZT remains unchanged in our experiments.

Graphene flakes are mechanically exfoliated on PZT followed by optical identification. Thin sheets are examined with atomic force microscopy (AFM) height measurements (Fig. 1a) and Raman spectroscopy (Fig. 1b). Selected flakes with height less than 5 nm ($n \leq 15$) are fabricated into Hall-bar field effect transistor (FET) devices using e-beam lithography and metal evaporation [16]. The doped STO substrates are conducting and serve as the back gate electrodes of the FET devices (Fig. 3a).

HfO$_2$–graphene–SiO$_2$ double-gate devices are fabricated by patterning and depositing high quality HfO$_2$ films on conventional SiO$_2$–gated graphene FETs using atomic layer deposition [27] (Figs. 2a and 2c). The HfO$_2$ films are amorphous and smooth, growing continuously across the graphene/SiO$_2$ step, as shown in Fig. 2b. Raman spectra obtained on both the bare and the HfO$_2$–covered sides of the graphene sheet are comparable to pristine graphene, with no visible D peak [27].

Resistance and Hall measurements are performed on graphene FETs with Hall bar configurations. Saturation current experiments are carried out on two terminal devices to ensure uniform current density. The dielectric constants $\kappa$ of PZT and HfO$_2$ films are deduced from both the gate-dependent Hall measurements and low frequency capacitance measurements [16, 27]. For the PZT films, $\kappa$ varies from 13-100. The dielectric constant of HfO$_2$ is 17±0.2 [27]. Both materials allow us to achieve larger than 1.5x10$^{13}$/cm$^2$ carrier density modulation in graphene, exceeding the range of the 300 nm SiO$_2$ back gate (~1x10$^{13}$/cm$^2$ at 140 V). The high dielectric constants of PZT and HfO$_2$ enable large, efficient charge modulation and dielectric screening in graphene transistors.

## 3. Graphene on Single Crystal Ferroelectric PZT
### 3.1 High Mobility and Signatures of Band Crossing and LA Phonon Scattering

Graphene FETs prepared on single crystal PZT substrates exhibit significantly improved carrier mobility, which allows us to observe features that are not accessible on SiO$_2$-gated graphene, such as clear signatures of band crossing in semi-metallic $n$-LG and the small resistivity due to the scattering of LA phonons in graphene. Figure 3c shows the sheet resistivity $\rho$ of a 7-LG device as a function of gate voltage $V_g$ at 4 K. At this thickness the *n*-LG behaves as a 2D semimetal with small overlap between the electron and hole bands (Fig. 3b) [35]. The carrier density is controlled by $V_g$ through $\Delta(n_e - n_h) = \alpha V_g$ in the band overlap regime (regime I) and $\Delta n_{e,h} = \alpha V_g$ in the pure 2D electron (regime II) and hole regimes, where $\alpha$ is the charge injection rate of the back gate.

In Fig. 3c, we observe a prominent kink in $\rho(V_g)$ at $V^T_g = 1.1$ V. This feature is smoothed out at $T > 10$ K [16]. This change of slope coincides with the system transitioning from the band overlap two-carrier regime (I) to the single-carrier electron regime (II). The signature of such band crossing has not been reported in $SiO_2$-gated *n*-LG [35], and is a clear indication of the high quality of the PZT-gated samples.

In the single-carrier regime, the appearance of a linear $\rho(T)$ due to LA phonon scattering is very pronounced, as a result of the small residual resistivity (Fig. 4a). Above the Bloch - Gruneisen temperature $T_{BG} = \hbar q v_{ph}/k_B$, where the phonon wavevector $q = 2k_F$ and $k_F$ is the electron Fermi wavevector, the resistivity due to LA phonon scattering is given by:

$$\rho_{LA}(T,n) = \frac{m_e^*}{ne^2}\left\langle\frac{1}{\tau}\right\rangle = \frac{1}{n}\frac{(m_e^*)^2 D^2 k_B T}{4\hbar^3 e^2 \rho_m v_{ph}^2} \quad \text{for } n\text{-LG with parabolic bands} \quad (1a),$$

and $\quad \rho_{LA}(T) = (h/e^2)\pi^2 D^2 k_B T/(2h^2 \rho_s v_s^2 v_F^2) \quad$ for 1-LG with linear bands $\quad$ (1b).

Here $D$ is the acoustic deformation potential, $\rho_m$ is the areal mass density of graphene, and $v_{ph}$ is the velocity of sound in graphene. This linear $T$-dependence is observed in $\rho(T)$ at all densities in the single-carrier regime II (Fig. 4a) above $T_{BG}$. In $SiO_2$-gated graphene, this linear $\rho(T)$ is superimposed on a large residual resistivity and superseded by a more rapid rise above ~150 K, which make an unambiguous identification difficult [12].

A fit to Eq. 1a yields a deformation potential $D = 7.8 \pm 0.5$ eV. Equations 1a and 1b, however, do not consider the dielectric screening of the electron-phonon interaction, which is likely to be strong in PZT-supported samples. Neglecting the dielectric screening leads to an underestimate of $D$, as demonstrated in GaAs 2D systems [36]. Further analysis adding a static Debye screening function in the random-phase approximation $|S|^2 = 1/(1+P/q)^2$ to the fit [37], where $P = (e^2/\kappa\varepsilon_0)(2m^*/\hbar^2)$, produces an upper bound of the deformation potential $D_{max} = 24$ eV, which is in line with results obtained on $SiO_2$-gated graphene (18 eV) [12, 27] and suspended graphene (29 eV) [15].

As $T$ drops below $T_{BG}$, large angle scattering events become increasingly suppressed and the phonon resistance vanishes more rapidly with temperature, following a $T^4$ power law [11, 38]. This regime is inaccessible in our devices as a $T$-independent resistance term $\rho_0(n)$ dominates below $T_{BG} \sim 80$ K, but has been experimentally confirmed in electrolyte-gated graphene where $T_{BG}$ was tuned to up to 1000 K [38].

The small $\rho_0$ corresponds to high low-temperature mobility $\mu > 1 \times 10^5$ cm$^2$/Vs. Figure 4b compares $\mu(T)$ obtained from a 7-LG device on PZT, a SiO$_2$-gated 7-LG, a SiO$_2$-gated single-layer graphene, bulk graphite from Ref. [39] and the intrinsic LA phonon-limited mobility calculated from Eq. 1a. In the single-carrier regime with $n_e = 2.4 \times 10^{12}$/cm$^2$, PZT-gated devices show $\mu \sim 7 \times 10^4$ cm$^2$/Vs at room temperature, close to the intrinsic phonon mobility of $\sim 1 \times 10^5$ cm$^2$/Vs. At low $T$, mobility up to $1.4 \times 10^5$ cm$^2$/Vs is observed, which is about one order of magnitude higher than those on SiO$_2$-gated graphene [12, 13].

PZT-gated $n$-LG ($n$ = 1-15) devices exhibit a large range of mobility values, varying from 16,000-140,000 cm$^2$/Vs. This variation appears to be correlated with two other characteristics of the films. First, the dielectric constant of the PZT films also varies considerably from $\kappa$ = 13 to 100 and the high mobility samples are always found on PZT films with large dielectric constant. Second, the degree of screening of the film's polarization by surface adsorbates appears to play a role. Although the PZT substrates possess a large spontaneous polarization, the polarization is almost completely screened under ambient conditions. As a result, the screening charges are trapped at the graphene/PZT interface upon exfoliation, and most of the devices only show a very small initial doping (less than 1% of the polarization) at room temperature. We find that lower mobility devices tend to show higher initial doping, suggesting a higher degree of incomplete screening. Overall, the high mobility in these samples in the presence of the high density interface charges suggest weak charged impurity scattering, which is probably due to the strong dielectric screening of PZT. It is also possible that the adsorbate layer may possess a high degree of order in registry to the lattice sites of the crystalline PZT substrate, and scattering is suppressed because of the ordering. Controlled studies show that on epitaxial PbTiO$_3$ surface, OH$^-$ chemisorbs on the Pb$^{2+}$ sublattice to screen its polarization, with a binding energy of ~200 meV [40].

**3.2 Unusual Resistance Hysteresis due to Dynamic Interfacial Adsorbate Screening**

The likely candidates that screen the polarization of the PZT film are charged adsorbates, including free ions, atoms and molecules in the ambient and OH$^-$ and H$^+$ produced by the dissociation of H$_2$O [40-42]. These screening molecules may respond dynamically to the change of PZT's polarization, which will be reflected in the gate-dependent resistance of graphene. Indeed, unusual resistance hysteresis has been observed on PZT-gated $n$-LG devices at high gate voltage. Figure 5a shows $\rho(V_g)$ of the 7-LG device at room temperature, which exhibits distinct behaviors at low and high $V_g$. Below 2 V, we observe the conventional field effect modulation, with the forward and backward gate sweeps producing similar $\rho(V_g)$ as well as $n(V_g)$. At $V_g > 2$ V, $\rho(V_g)$ becomes hysteretic, with the backward sweeping curve shifted to the right of the forward sweeping curve by a density level of $\Delta n = 2.7 \times 10^{12}/cm^2$. This shift corresponds to only one percent of PZT's polarization ($\sim 3 \times 10^{14}/cm^2$). Similar hysteresis is observed on several PZT-gated n-LG devices fabricated on different PZT films, regardless of the thickness of the graphene sheet (2-15 layers), its carrier mobility (16,000-140,000 $cm^2/Vs$) and the dielectric constant (30-100) of the PZT (Fig. 5b).

The hysteresis is clearly not induced by ferroelectric switching, as its direction is opposite to that expected from the density modulation induced by PZT's polarization reversal. The onset of this "anti-hysteresis" is accompanied by saturation in $\rho$ and $n$, and the onset voltage is smaller than the coercive voltage of PZT.

In Figs. 5a and 5b, the black curves are the more stable state at lower $V_g$, while the red curves are more stable at higher $V_g$. The relaxation between the meta-stable states in $\rho(V_g)$ is characterized by an exponential time dependence $\exp(-t/\tau)$ with the relaxation time constant $\tau$ increasing from 6 hours at 300 K to 80 days at 77 K [17]. The $T$-dependence of the relaxation rate can be well described by a thermally activated relaxation process, $\frac{1}{\tau} \sim \exp\left[-\frac{\Delta E_b}{k_B T}\right]$, with an activation barrier $\Delta E_b$ of 50-110 meV. This is a robust phenomenon that may be potentially useful in constructing graphene non-volatile memories if one can engineer the relaxation time to be longer.

Such anti-hysteresis behavior has also been observed in carbon nanotube FETs gated by $SiO_2$ and ferroelectric $BaTiO_3$ [43-45]. A plausible scenario involves the dynamical screening of PZT's polarization through the dissociation-recombination of interfacial water molecules. Water chemisorbed on the surface of transition metal oxides are known to have two metastable forms: the molecular form $H_2O$ and the dissociated state as $H^+$ and $OH^-$ [41, 42]. The balance between the dissociation and recombination processes is influenced by the geometry of the lattice, the presence of defects or uncompensated charges, and external electric fields [41].

As likely candidates that screen PZT's polarization under ambient condition, large densities of dissociated $OH^-$ and $H^+$ and chemisorbed water are expected to be trapped at the PZT/graphene interface. The as-grown PZT films are uniformly polarized downward (Fig. 3a). When a positive $V_g$ is applied, the polarization field decreases, which results in over-screening from the adsorbates and favors water recombination to decrease the density of screening charge. Conversely, as $V_g$ decreases, $P$ increases and results in under-screening from the adsorbates. Dissociation of water is then favored to provide additional screening. Both processes are thermally activated with an activation barrier on the order of the binding energy between $H^+$ ($OH^-$) and the oxide surface $O^{2-}$ ($Pb^{2+}$). The estimated barrier height $\Delta E_b \sim$ 50-110 meV is consistent with the binding energy of $H^+$ and $OH^-$ on transition metal oxide surfaces [40, 41]. As both processes attempt to compensate PZT's polarization change, they screen the field effect modulation from graphene and lead to the observed saturation in resistance and carrier density and the anti-hysteresis. As this process is solely determined by the interface chemistry between water and PZT, remarkable similarity in the anti-hysteresis has been observed among devices with different graphene layers and dielectric properties of PZT.

Recently, Zheng *et al*. have successfully demonstrated resistance hysteresis due to ferroelectric switching on graphene using ferroelectric poly(vinylidene fluoride-trifluoroethylene) as the top gate [21, 22]. Compared to polymeric materials, oxides have the distinctive advantage of higher switching speed, robust mechanical properties, and lower switching voltage. However, a controlled graphene-oxide interface is necessary to implement such device concepts.

## 4. Effect of Remote Surface Optical Phonon of the Oxide Layer
### 4.1 Effect of RSO Phonon on High Temperature Mobility

Scattering from remote polar optical phonon in neighboring oxide layer is known to be one of the major mobility-limiting factors in silicon transistors, especially those using high-$\kappa$ oxides with soft phonon modes [23]. Electrons in graphene are also subject to this mechanism at elevated temperatures [46]. In graphene the temperature dependent resistivity $\rho(T)$ is affected by three contributions:

$$\rho(T, n) = \rho_0(n) + \rho_{LA}(T) + \rho_{RSO}(T, n) \qquad (2),$$

where $\rho_0(n)$ is the $T$-independent residual resistivity, and $\rho_{LA}(T)$ and $\rho_{RSO}(n, T)$ are from the LA phonon and RSO phonon contributions, respectively. The last term is given by:

$$\rho_{RSO}(T,n) = \int A(\mathbf{k},\mathbf{q}) d\mathbf{k} d\mathbf{q} \sum_i g_i / \left(e^{\hbar \omega_i / k_B T} - 1\right) \qquad (3),$$

where $A(\mathbf{k}, \mathbf{q})$ is the matrix element for scattering between electron ($\mathbf{k}$) and phonon ($\mathbf{q}$) states, and $\omega_i$ and $g_i$ represent the frequency and coupling strength, respectively, of the $i$th surface optical phonon mode [23, 46, 47].

The effect of RSO phonons on $SiO_2$-gated graphene was first examined by Chen *et al*. [12]. Two RSO phonon modes in $SiO_2$ with $\omega_1 = 59$ meV and $\omega_2 = 155$ meV are shown to limit graphene's room temperature mobility to 40,000 cm$^2$/Vs [12, 47]. We employ a double-oxide $HfO_2$-graphene-$SiO_2$ structure to investigate how RSO phonon scattering affects the electron mobility in graphene. Both oxides contribute RSO phonon modes, and the strength of the modes is determined by the dielectric environment, which is different from the single-oxide situation.

For a quantitative comparison, $\rho(T)$ of graphene FETs has been obtained from $SiO_2$-only-gate and double-oxide-gate devices fabricated on the same graphene sheet (Fig. 2a).

The high quality of the HfO$_2$ deposition enables a low temperature field effect mobility $\mu_{FE}$ as high as ~20,000 cm$^2$/Vs in HfO$_2$-covered graphene, comparable to the best pristine exfoliated graphene on substrates [27]. Figure 6a shows $\rho(T)$ taken on one of these samples on the bare and covered side at $n = 3\times10^{12}$/cm$^2$. Below 100 K, both $\rho(T)$ are well described by the linearly $T$-dependent LA phonon scattering, and the slope 0.1 Ω/K corresponds to a deformation potential $D_A$=18±2 eV, in good agreement with the value reported in literature [12].

Above 100 K, both $\rho(T)$ increase supralinearly with $T$, with the HfO$_2$-covered side exhibiting a much steeper rise in $\rho$ in all samples. The temperature dependences can be well described by the RSO model using Eqs. 2 and 3 (Fig. 6a). On the vacuum-graphene-SiO$_2$ side of the device, two RSO phonon modes from the SiO$_2$ substrate are important: $\omega_1$ = 63 meV, $\omega_2$ = 149 meV, $g_1$ = 3.2 meV, and $g_2$ = 8.7 meV [12, 47]. The phonon distribution of amorphous HfO$_2$ are approximated with a single frequency $\omega_3'$ = 54 meV. In the double-oxide geometry, the presence of an additional oxide layer modifies the intrinsic surface phonon modes slightly and screens their coupling strength to satisfy the boundary condition $\varepsilon_{SiO_2}(\omega) + \varepsilon_{HfO_2}(\omega) = 0$. We obtain the modified phonon modes to be $\omega_1'$ = 72 meV, $\omega_2'$ = 143, meV, $g_1'$ = 1.2 meV, and $g_2'$ = 2.4 meV for SiO$_2$, and $\omega_3'$ = 54 meV and $g_3'$ = 5.7 meV for HfO$_2$ [27]. Details of how these phonon modes are calculated can be found in Refs. [23, 27].

An alternate proposal employs the thermal activation of quenched ripples to explain the rapid rise in $\rho(T)$ above 100 K in SiO$_2$-gated devices [13]. This model is hard to reconcile with our results obtained on the double-oxide devices, as the presence of the HfO$_2$ overlayer clearly introduces additional scattering channels.

Figure 6b compares mobilities $\mu_i(n)=1/ne\rho_i$ determined for various phonon channels $i$ in a typical device [27]. The LA phonon limit of mobility scales as ~1/$n$, which is approximately 1x10$^5$ cm$^2$/Vs at $n = 2\times10^{12}$/cm$^2$ and 300 K. The $\omega_1$ mode of the SiO$_2$ substrate limits room temperature $\mu$ to ~60,000 cm$^2$/Vs in single-oxide devices. Its contribution is significantly suppressed in HfO$_2$-graphene-SiO$_2$ devices due to screening from the HfO$_2$ overlayer, imposing a high mobility limit of $\mu$ ~ 2x10$^5$ cm$^2$/Vs. In the double-oxide structure, the RSO phonon modes of HfO$_2$ dominate scattering and limit $\mu$ to approximately 20,000 cm$^2$/Vs at 300 K. These results reveal the impact of the soft phonon modes of high-$\kappa$ oxides on carrier mobility in graphene, which needs to be taken into consideration in the design of graphene-high-$\kappa$ oxides hybrid electronics.

**4.2 Saturation Current at High Source-Drain Bias in Single- and Double-Oxide Gated Devices**

In this section we discuss the effect of RSO phonons on transport in graphene FETs operating in the high source-drain bias $V_{sd}$ regime. At low bias $V_{sd}$, the current density in graphene is described by the Drude model with $j = neu = ne\mu E$ ($E = V_{sd}/L$ is the transverse electric field and $u$ is the drift velocity), and the drain current $I_d$ is linearly proportional to $V_{sd}$. The drift velocity and current reach saturation at large $V_{sd}$ due to increasing energy lost in inelastic scattering with phonons. Recent high-field transport measurements have reported highly non-linear high field $I(V_{sd})$ (*I-V*) and a remarkable current density $j$ of a few mA/μm in graphene transistors [25, 48-51], comparable to that of carbon nanotubes (CNTs) [52, 53] and exceeding the performance of silicon transistors [9]. A high saturation velocity/current and a low output conductance in the saturated regime make graphene FETs promising for high-frequency linear amplifiers.

The current saturation in CNTs can be well described by $I = \dfrac{V}{R_0 + V/I_{sat}}$, where $R_0$ is the small-bias resistance of the sample and $I_{sat}$ the saturation current [52, 53]. This model assumes instantaneous emission of optical phonon by the electrons that are accelerated in an electric field, i.e. the electron-phonon backscattering length $l_{ph}$ is much shorter than the acceleration length for the electron to reach the phonon energy $\hbar\Omega$, $l_\Omega = \hbar\Omega/eE$, a condition that is satisfied in CNT. In graphene, since $l_{ph}$ is not as short due to weaker electron-phonon coupling [26, 48], the instantaneous emission model does not apply. It has been shown that the current in graphene does not reach full saturation [26, 49] except in the presence of a carrier density gradient [25], which has been suggested to be due to the competition between disorder and phonon scattering.

The current saturation in graphene can be well understood within a full microscopic theory that takes into account the effects of impurity and phonon scattering. Considering the contributions from all scattering channels, the high-field transport in graphene can be described by the Boltzmann equation:

$$-\frac{eE}{\hbar} \cdot \nabla f_{k\alpha} = S_{col} = S_{col}^{imp} + S_{col}^{imp'} + S_{col}^{LA} + S_{col}^{LO} + S_{col}^{RSO} \qquad (4).$$

The scattering integral $S_{col}$ on the right originates from impurities and phonons, including the contributions from charged impurities (imp), neutral scatterers (imp') [12, 29], the LA [11, 12] and longitudinal optical (LO) phonons of graphene [48, 51], and the RSO phonons of the substrate [12, 46, 47]. The phonon terms have the form

$$S_{col} = -\sum_{\mathbf{p}\gamma} \left[ f_{\mathbf{k}\alpha}(1 - f_{\mathbf{p}\gamma}) W_{\mathbf{kp}}^{\alpha\gamma} - f_{\mathbf{p}\gamma}(1 - f_{\mathbf{k}\alpha}) W_{\mathbf{kp}}^{\gamma\alpha} \right]$$

where $W_{\mathbf{kp}}^{\alpha\gamma} = (2\pi/\hbar) \sum_{\mathbf{q}s} \delta_{\mathbf{q+k-p}} |M_{\mathbf{kp}}^{\alpha\gamma}|^2 \left(N_q + \dfrac{1}{2} - \dfrac{s}{2}\right) \delta(\varepsilon_{k\alpha} - \varepsilon_{p\gamma} + s\hbar\omega_q)$ (5).

Here $s = +1$ (-1) for phonon absorption (emission) and $M_{\mathbf{kp}}$ is the matrix element for electron scattering from momentum state **k** in band $\alpha$ to state **p** in band $\gamma$. $N_q$ is the phonon occupation factor, and $\hbar\omega_q$ is the phonon energy. Here the phonon energy is assumed to be in equilibrium with the graphene lattice and the substrate, and the electron-electron interaction is included implicitly in the "displaced" Fermi-Dirac distribution:

$$f_{\mathbf{k}\alpha} = \left[\exp\left(\varepsilon_{k\alpha} - \hbar\mathbf{u}\cdot\mathbf{k} - \mu_e\right)/k_B T_e + 1\right]^{-1} \qquad (6).$$

Here $\alpha = \pm 1$ denotes the conduction or valance band, $\varepsilon_{k\alpha} = \alpha\hbar v_F k$ is the energy spectrum of graphene, and $\mu_e$ is the chemical potential. It assumes that the electron-electron scattering time is sufficiently short that electrons come to equilibrium before any other scattering processes occur [54].

We have studied the low temperature $I$-$V$ on $SiO_2$-gated and $HfO_2$-graphene-$SiO_2$ two terminal devices. The scattering terms from impurities $S_{col}^{imp}$ and $S_{col}^{imp'}$ are extracted by fitting to $\sigma(V_g)$ measured on the four terminal devices fabricated on the same graphene sheet [26]. Figure 7a compares the measured drift velocity as a function of electric field with various models. The most striking result is that the full theory as well as the model considering only impurities and RSO phonons give an excellent description of the measured $I$-$V$ in graphene with no adjustable parameters in our calculations [26]. RSO phonons are the principal scattering mechanism for high-field transport, and more than 95% of the power dissipation occurs through the RSO phonons. A model that only considers LA and LO phonons of graphene results in a substantial underestimate of $u_{sat}(E)$. It should be noted that the inclusion of RSO phonons leads to an increase of the electron drift velocity. This is because they provide an efficient route for energy dissipation, leading to a drastic drop of electron temperatures and higher saturation velocities.

Given the crucial role played by the RSO phonons, we have modeled the saturation current in graphene devices on several commonly used high-$\kappa$ substrates, including $HfO_2$, $Al_2O_3$, and $ZrO_2$ [26]. Figure 7b shows the calculated $u(E)$ for various oxides in the vacuum-graphene-oxide structures with different RSO phonon energies and coupling constants taken from Ref. [23] ($ZrO_2$, $Al_2O_3$) or our measurements [27] ($SiO_2$ and $HfO_2$). Despite the large variation of $\omega_{RSO}$, the saturated velocity values differ by less than 25%. This behavior is partly due to the trend of decreasing coupling strength with decreasing $\omega_{RSO}$ and partly due to the competing effect of electron cooling and momentum scattering. The electron drift velocity $u(E)$ is the highest on $Al_2O_3$, followed by $SiO_2$, $HfO_2$ and $ZrO_2$. The onset of saturation follows the opposite order.

We have also studied double-oxide $HfO_2$-graphene-$SiO_2$ FETs. As shown in Fig. 7c, the full theory and the theory considering only the RSO phonons of $HfO_2$ both provide excellent agreement (within 4%) with the experimental data. The saturated drift velocity in the double-oxide structure, surprisingly, is 10% lower than that for single-oxide devices on $SiO_2$ substrates and 4% lower than that for devices on $HfO_2$ substrates [26]. This may be due to a less efficient cooling of the electrons through RSO phonons due to the enhanced dielectric screening.

Our calculations show that current saturation value displays a linear *n*-dependence at high carrier densities, with the saturated current density in the range of a few mA/μm [26]. This value translates into a bulk current density exceeding $10^8$ A/cm$^2$, which is extraordinary and advantageous in amplifier applications. The density of Coulomb impurities has little effect on $u_{sat}$, although the velocity saturation occurs at lower bias in cleaner samples [26, 55]. These predictions can be used to guide the design and optimization of high-frequency graphene linear amplifiers.

## 5. Quantum Scattering Time – Charged Impurity Scattering and Dielectric Screening

At low temperature, all phonon contributions to resistivity in graphene diminish. The origin of major scattering sources that impose the low temperature mobility limit in graphene devices prepared on substrates is a central issue that is highly debated [29, 56]. It has been shown that charged impurity (CI) is one of the major extrinsic scattering sources that limit mobility [29, 57]. Other candidates include ripple scattering [58] and resonant scatterers [59, 60]. Within the CI model, the conductivity of graphene has been explained by combining scattering from long- and short-ranged sources, and the presence of dielectric screening is expected to reduce the Coulomb potential of charged impurities and enhance mobility [61-63]. Controversial experimental results have been reported regarding the effect of dielectric screening. Improvement of mobility has been found in PZT-gated graphene [16] and graphene with top dielectric layers such as ice [64] and solvents [65], but certain liquid dielectric layers produce screening effects much smaller than expected from the CI model [66]. The surface properties of the dielectric layers also appear to play an important role, as high mobility graphene-on-substrate devices have only been realized on single crystal substrates such as PZT [16] and BN [18], with the dielectric constant of the latter only comparable with that of $SiO_2$. Within the CI model, the origin of such impurities remains unclear: adsorbates on top of graphene, charges adsorbed/trapped at the graphene/oxide interface or residing inside the neighboring dielectrics are all possible candidates.

A widely employed parameter to evaluate the effect of electron scattering is the carrier mobility $\mu$, or equivalently the transport scattering time $\tau_t = m^*\mu/e$. [29, 56]. The transport scattering time, however, is only sensitive to scattering events with large scattering angles and does not yield enough information to differentiate various types of scattering sources. Critical information can be obtained by evaluating another important parameter, the quantum scattering time $\tau_q$ [33, 34]. $\tau_q$ characterizes the momentum relaxation of a quasi-particle with corresponding quantum level broadening $\Gamma=\hbar/2\tau_q$. Quantitatively, $\tau_q$ and $\tau_t$ in graphene are given by the following equations [32]:

$$\frac{1}{\tau_q} = n_{imp} \int_0^\pi Q(\theta, k_F)(1+\cos\theta) d\theta$$
$$\frac{1}{\tau_t} = n_{imp} \int_0^\pi Q(\theta, k_F)(1+\cos\theta)(1-\cos\theta) d\theta \qquad (7),$$

where $\theta = \theta_{kk'}$ is the angle between the initial and final wave vector **k** and **k**', $n_{imp}$ is the impurity density and $Q(\theta, k_F)$ depends on specific scattering mechanisms [32, 67]. The factor $(1+\cos\theta)$ results from the unique pseudo-spin conservation in graphene, which suppresses 180° backscattering. The factor $(1-\cos\theta)$ reduces the impact of small-angle scattering on $\tau_t$. As a result, $\tau_t$ is mostly affected by right angle scatterings, while $\tau_q$ is heavily affected by small-angle events. Measurement of $\tau_t / \tau_q$ can thus provide critical information about the dominating scattering scenarios in conventional 2D electron gases [67-69]. For example, in modulation-doped GaAs/Al$_x$Ga$_{1-x}$As heterostructures, $\tau_t/\tau_q$ routinely reaches values as high as several multiples of ten, since the ionized donors are far away from the conducting channel [67-69]. This ratio is close to 1 in silicon inversion layers, since both interfacial charges and surface roughness contribute roughly equally to $\tau_t$ and $\tau_q$ [67, 69].

Quantitative comparison between $\tau_t$ and $\tau_q$ in graphene has been examined experimentally in Refs. [33] and [34] to evaluate different scattering scenarios. In both works $\tau_q$ is extracted from the magnetic field dependence of Shubnikov-de Haas (SdH) oscillations, also called the Dingle plot (Figs. 8a and 8b), which is described by:

$$\frac{\delta\rho}{\rho_0} = 4\gamma_{th} \exp\left(-\frac{\pi}{\omega_c \tau_q}\right); \quad \gamma_{th} = \frac{2\pi^2 k_B T/\hbar\omega_c}{\sinh(2\pi^2 k_B T/\hbar\omega_c)} \qquad (8).$$

Here $\rho_0$ is the non-oscillatory background resistance, $\delta\rho_{xx}$ is the oscillatory amplitude, $\gamma_{th}$ is the thermal factor, and $\omega_c = eB/m^*$ is the cyclotron frequency with $m^*$ the effective mass of graphene. This approach has also been used to determine the effective mass of carriers in bilayer graphene [70].

We have carried out comprehensive studies of $\tau_t$ and $\tau_q$ in graphene samples prepared on SiO$_2$ and PZT substrates and HfO$_2$-graphene-SiO$_2$ structures. These samples exhibit a wide range of mobility 4,400 < $\mu$ < 22,000 cm$^2$/Vs. $\tau_q$ ranges approximately 25-120 fs in these samples, corresponding to $\gamma$ = 3-13 meV [33]. Despite the different dielectric materials and large variation in the values of $\tau_t$ and $\tau_q$, the magnitude and $n$-dependence of the ratio $\tau_t/\tau_q$ can both be described by the CI model [29, 32] by decomposing the long- and short-ranged contributions and varying the CI-graphene distance $z$.

In SiO$_2$-gated samples, there is a large variation in $\tau_t/\tau_q$ ranging from 1.7 to 5.4 and no direct correlation between mobility and $\tau_q$ values, further confirming that $\tau_t$ alone does not give a complete description of relevant scattering sources in graphene [33]. Figure 8c plots the long-ranged component $\tau_t^{long}/\tau_q^{long}(n)$ obtained on five SiO$_2$-gated graphene samples. The long- and short-ranged components in $\tau_{t,q}$ are separated using the following equations [32]:

$$\frac{1}{\tau_{t,q}} = \frac{1}{\tau_{t,q}^{long}} + \frac{1}{\tau_{t,q}^{short}};$$
$$\tau_t^{short} = \frac{m^*}{ne^2 \rho_{short}}; \quad \frac{\tau_t^{short}}{\tau_q^{short}} = s \quad (9).$$

Here $\rho_{short}$ is the resistivity from short-ranged scatterers, and $s$ is a constant determined by the dielectric environment. For the SiO$_2$/vacuum geometry, $s = 1.1$ [32]. The $n$-dependence of $\tau_t^{long}/\tau_q^{long}$ falls into two groups. The first group of samples shows $\tau_t^{long}/\tau_q^{long} \sim 2$ that is roughly $n$-independent. The second group of samples show ratios larger than 2, which increase with increasing $n$. For the dielectric environment of vacuum/SiO$_2$, the CI model predicts an $n$-independent constant $\tau_t^{long}/\tau_q^{long}(n) \sim 2.5$ for charged impurities located right at the graphene plane ($z = 0$) [32], and a ratio increasing with increasing $n$ for a finite $z$ due to enhanced screening. As shown in Fig. 8c, both groups of behavior can be well described by CIs located within 0 to 2 nm of the graphene sheet.

As $\tau_t/\tau_q$ is also affected by screening, it can also be tuned by the dielectric constant of oxides. We have studied $\tau_t/\tau_q$ on HfO$_2$ ($\kappa = 17$)-graphene-SiO$_2$ and PZT ($\kappa = 13$)-gated graphene devices (Fig. 8d), where $s = 1.26$ and 1.2, respectively. For both devices, $\tau_t^{long}/\tau_q^{long}$ results are again well described by the CI model and point to interfacial charged impurities ($z = 0$) as the most important source of scattering.

In the literature, various scenarios have been proposed to account for the mobility ceiling observed in substrate-supported graphene [29, 32, 56, 71]. Atomic defects, for example, induce midgap state scattering that produces a $\sigma(n)$ similar in shape to that caused by charged impurities [56, 60]. In this scenario, $\tau_t/\tau_q$ is a constant between 1 and 2, independent of $n$ and dielectric constant [34]. In Ref. [34], Monteverde $et$ $al.$ report an $n$-independent $\tau_t/\tau_q \sim 1.7$ and interpreted their results as evidence for midgap state scattering. While such a ratio is consistent with both midgap state scattering and scattering from CIs at $z = 0$ combined with short-ranged scatterers, $\tau_t/\tau_q$ ratios greater than 2, and their density and dielectric environment dependence revealed in our studies are only well explained by the CI model. Our study of $\tau_t/\tau_q$ thus provides critical information in differentiating various scattering scenarios in graphene.

Fitting $\tau_t$ and $\tau_q$ to Eq. 7 reveals that $n_{imp}$ ranges from $3\times10^{11}/cm^2$ to $1\times10^{12}/cm^2$ in our samples [33]. Since the dominant CIs reside within 2 nm of the graphene sheet, primary candidates include charged adsorbates at the graphene surface or the graphene/$SiO_2$ interface, or oxide charges in bulk $SiO_2$. The oxide charges in current MOSFETs are generally in the low $10^{11}/cm^2$ regime, the majority of which are present at the Si-$SiO_2$ interface, not the bulk of the oxide [72]. They are thus too small to account for the observed scattering times. As to surface contaminants and adsorbates, approaches such as ultra-high vacuum baking [73, 74] and current annealing [75] have been employed to clean $SiO_2$-gated graphene, with no significant improvement to mobility observed. These observations collectively point to interfacial charges trapped at graphene/$SiO_2$ interface as the primary scattering source for $SiO_2$-gated graphene.

Given the high mobility observed on graphene prepared on BN [18], it is clear that the surface chemistry of the dielectric layer plays an important role. In ambient conditions, the hydrated surface of silica is covered with silanol (Si-OH) groups, possibly multi-layers of water, and $H^+$ and $OH^-$ ions dissociated from water [76]. The $O_2$ dissolved in water further promotes complex surface redox chemistry [77]. In addition, molecular dynamics calculations have shown that the surface states of silica can transfer electrons to graphene, resulting in $n$-doping and CI scattering. This doping was indeed observed after prolonged pumping of devices in vacuum at high temperatures [78]. Although the details of the above aspects remain to be fully examined, it is likely that the surface charges of silica play a major role in limiting graphene's mobility, while uncontrolled multi-layers of water serve as a spacer. Our measurements of $\tau_t/\tau_q$ are consistent with this scenario.

## 6. Conclusion

In this paper, we have reviewed different aspects of integrating functional oxides with graphene. Significant performance improvement has been achieved in $n$-layer graphene FETs by interfacing them with single crystal ferroelectric gate oxide PZT. The high-$\kappa$ nature of PZT (up to 100) and the superb room temperature carrier mobility in PZT-gated $n$-LG devices (up to 70,000 $cm^2$/Vs) make them promising for building electronic applications operating at small gate voltages and exceedingly high frequencies. Robust resistance hysteresis is observed in PZT-gated $n$-LG devices and attributed to the dynamics of the interfacial adsorbates. The understanding and control of the interface chemistry are critical for the future development of graphene-ferroelectric hybrid devices.

We have also evaluated the effect of remote surface optical phonons on the mobility and saturation current in graphene. $HfO_2$ top gate dielectrics limit carrier mobility in graphene to 20,000 cm$^2$/Vs at 300 K due to its soft remote surface optical phonon modes. By combining careful experimental and theoretical studies, we have demonstrated that at high source-drain bias, hot electrons in graphene lose energy predominantly by emitting the RSO phonons of the substrate. Our results provide valuable insights into the understanding of the electron transport in graphene in the presence of a dielectric layer and guide the design and performance optimization of high-speed graphene transistors.

We have systematically studied the quantum and transport scattering times in graphene interfaced with different oxides to critically examine different scattering scenarios. Our results can be quantitatively understood within the charged impurity model, from which we extract the density and location of charged impurities and assess the effect of dielectric screening in graphene. The experimental results of $\tau_t/\tau_q$ indicate that charged impurities residing within 2 nm of the graphene sheet are the main sources of scattering in graphene. Such information provides critical input towards the goal of eliminating these scattering sources and optimizing graphene's performance.

**Acknowledgements**

This work is supported by NSF Grants CAREER No. DMR-0748604 and MRSEC No. DMR-0820404. X.H. acknowledges support from NSF CAREER No. DMR-1148783 and University of Nebraska Research Council. C.H.A. acknowledges support from NSF DMR-1006256. The authors acknowledge use of facilities at the PSU site of NSF NNIN.

**Figure Captions:**

Fig. 1 a) AFM image of a few-layer graphene sheet on a 300 nm PZT film with rms roughness of 3-4 Å. The layer numbers are determined by height measurements. b) Raman spectra on $n$-LG normalized to the G peak intensity. The increasing background at low wave numbers and the small broad peak centered at 1615 cm$^{-1}$ are from the PZT substrate. Reproduced from Ref. [17]. Copyright (2010) by the American Institute of Physics.

Fig. 2 a) Optical image of a graphene FET that is partially covered by a HfO$_2$ overlayer. b) AFM of the regime circled in a) with a line cut measurement across the graphene/SiO$_2$ step. The rms roughness is 3–4 Å on graphene and 2–3 Å on SiO$_2$. Adapted from Ref. [27]. Copyright (2010) by The American Physical Society. c) SEM image of a HfO$_2$-graphene-SiO$_2$ FET. Adapted from Ref. [31]. Copyright (2010) by the American Physical Society.

Fig. 3 a) Schematic view of a $n$-LG FET gated by PZT. B) Schematics of the band structure of few layer graphene. $\delta\varepsilon$ ~30 meV for the 7-LG. c) $\rho(V_g)$ of a 7-LG at 4 K (open symbols). For this device the dielectric constant of PZT $\kappa$ is 100. The red solid fitting curve assumes a density-dependent mobility $\mu \sim n^\beta$ with $\beta = 0.9$ in the band overlap regime (I) and 1.3 in the electron-only regime (II). The dashed line is calculated assuming a density-independent mobility. The kink at $V_g^T = 1.1$ V (dash-dotted line) marks the transition between regimes I and II. Adapted from Ref. [16]. Copyright (2009) by the American Physical Society.

Fig. 4 a) $\rho(T)$ of a 7-LG at selected densities in the electron-only regime (II) with fits to the LA phonon model (Eq. 1a). b) Comparison of $\mu(T)$ in various graphitic materials. Solid squares: PZT-gated 7-LG at $n = 2.4\times10^{12}$/cm$^2$. Open triangles: a SiO$_2$-gated 7-LG at the same density. Open circles: single-layer graphene on SiO$_2$. Crosses: mobility of bulk graphite from Ref. [39]. Solid line: LA phonon-limited mobility calculated from Eq. 1a. Adapted from Ref. [16]. Copyright (2009) by the American Physical Society.

Fig. 5 Hysteresis loop in $\rho(V_g)$ taken on a) a 7-LG FET at 300 K and b) a 2-LG FET at 100 K. The 2-LG device has mobility of 16,000 cm$^2$/Vs and is fabricated on a PZT film with a dielectric constant of $\kappa \approx 30$. Arrows indicate the sweeping direction of $V_g$. Adapted from Ref. [17]. Copyright (2010) by the American Institute of Physics.

Fig. 6 a) $\rho(T)$ of a graphene FET partially covered by HfO$_2$ at $n = 3.0\times10^{12}$/cm$^2$ for the bare (black) and covered sides (red). Solid lines are fits to Eqs. 2 and 3. b) Mobility limit at 300 K imposed by scattering due to LA phonons in graphene and different RSO phonon modes in $SiO_2$ and $HfO_2$. Adapted from Ref. [27]. Copyright (2010) by the American Physical Society.

Fig. 7 a) The measured (circles) and calculated (lines) drift velocity (in units of Fermi velocity, $v_F$) vs. electric field. Dashed line: theory with only impurities and the LA and LO phonons of graphene. Dash-dotted line: theory with only impurities and the SO phonons of the $SiO_2$ substrate. Upper solid line: the full theory. Lower solid line: phenomenological model assuming instantaneous emission of the relevant optical phonons. Reproduced from Ref. [26]. Copyright (2010) by the American Physical Society. b) Calculated drift velocity for different substrates with $n = 1.9\times10^{12}/cm^2$ and $n_{imp} = 5\times10^{11}/cm^2$. c) Measured (blue circles) and calculated drift velocity for a $HfO_2$-graphene-$SiO_2$ device. Solid line: the full theory. Open squares: theory with only the unmodified $HfO_2$ RSO phonon of 75 meV. Solid squares: theory with only the 72 meV $SiO_2$ RSO phonon $\omega_1'$. Green dashed line: model assuming instantaneous emission of the RSO phonons of $HfO_2$. Red dash-dotted line: theory with LA and LO phonons. For all data in this plot, $n = 1.9 \times 10^{12}/cm^2$, $n_{imp} = 1.37\times10^{12}/cm^2$ and neutral impurities have no contribution to either theory or experiment. In the calculation, the substrate temperature $T_s = 20$ K, and the lattice temperature $T_L$ varies from 20 K to 260 K.

Fig. 8 a) $\rho(B)$ taken on a $SiO_2$-gated graphene device ($\mu_E =14,500$ cm$^2$/Vs) exhibits SdHO at low field and quantum Hall effect at high field. b) The corresponding Dingle plot reveals $\tau_q = 110$ fs. c) $\tau_t^{long}/\tau_q^{long}$ ($n$) for five $SiO_2$–gated graphene samples. The dashed lines are calculated ratios for short-ranged scatterers and long-ranged scatterers with various $z$. Adapted from Ref. [33]. Copyright (2009) by the American Physical Society. d) $\tau_t^{long}/\tau_q^{long}$ ($n$) for a PZT-gated graphene device (red diamonds) and a $HfO_2$-graphene-$SiO_2$ device (black squares). The dashed lines are calculated ratios for long-ranged scatters in different dielectric environments with $z = 0$.

# Figure 1

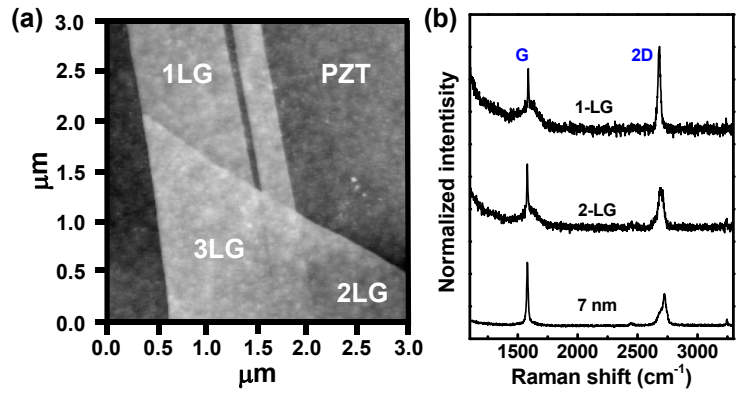

Figure 2

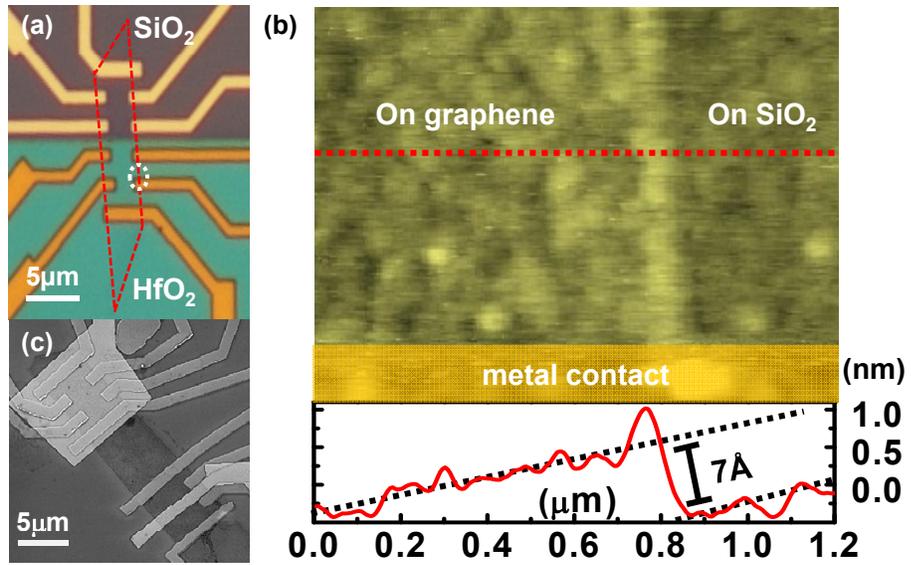

Figure 3

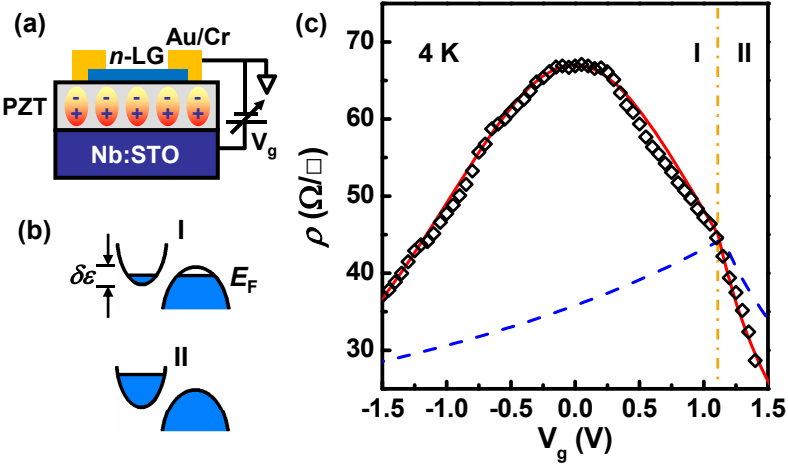

Figure 4

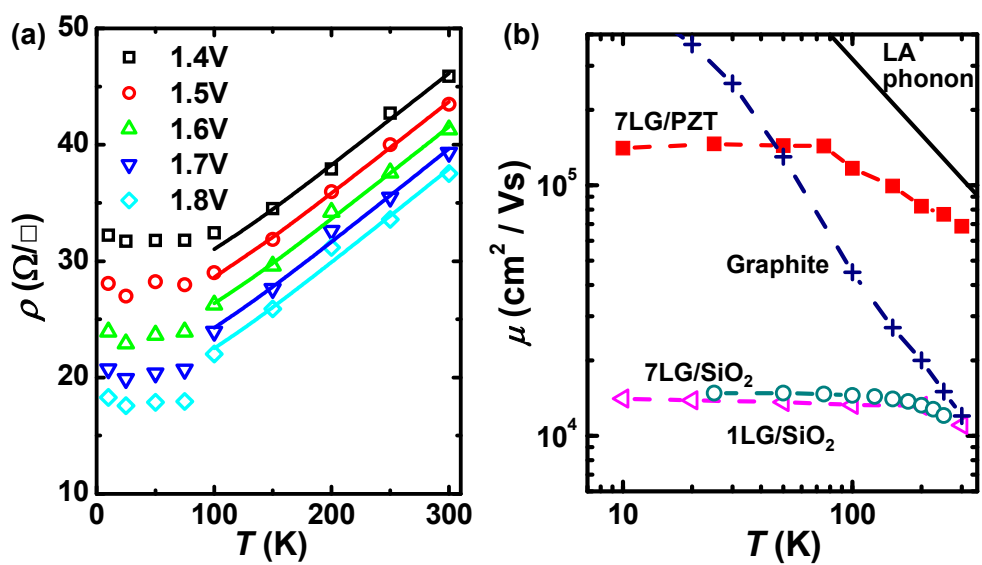

# Figure 5

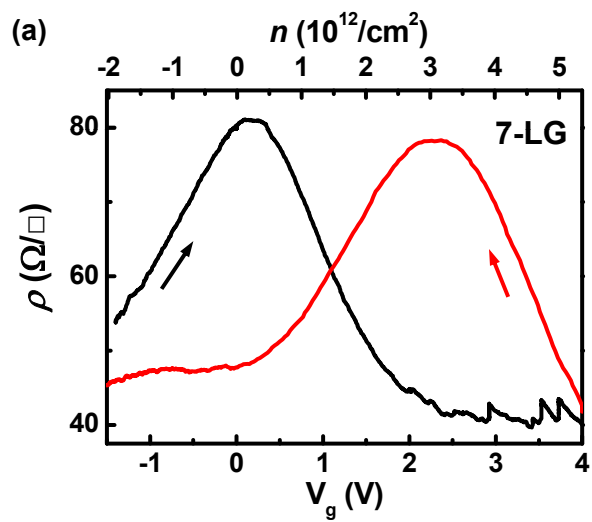

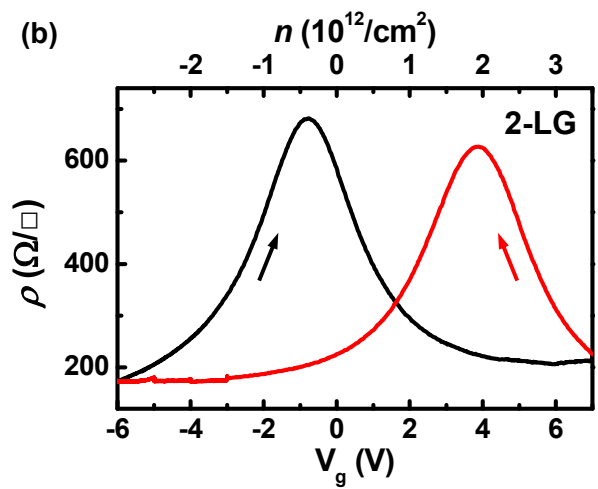

Figure 6

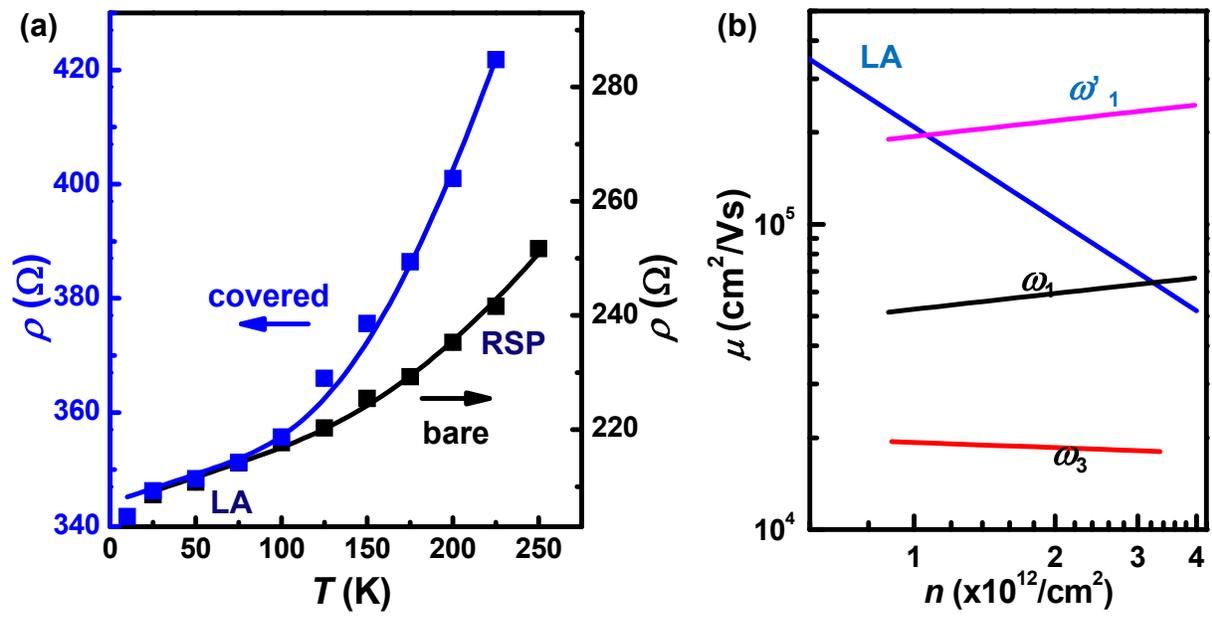

Figure 7

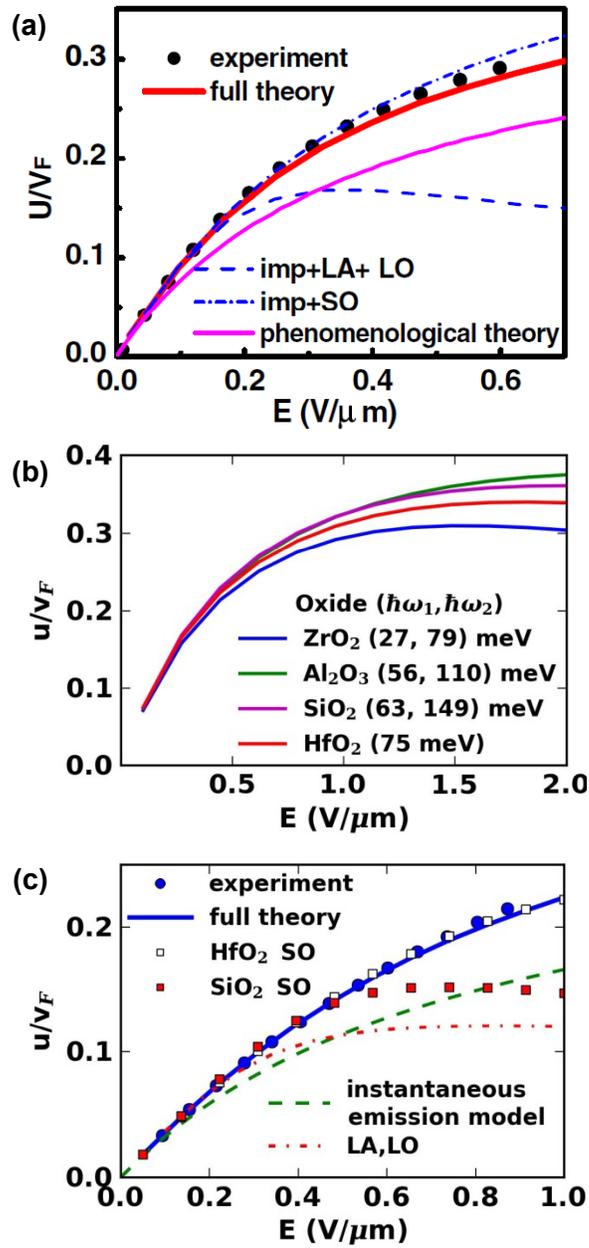

Figure 8

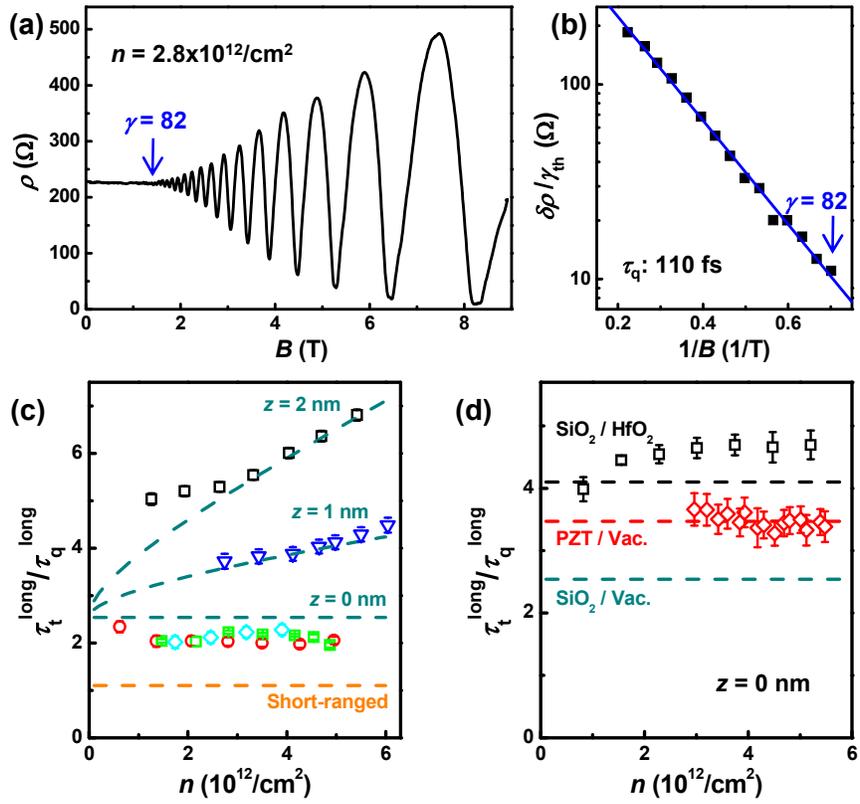